\documentclass[prl,
                reprint,
                amsmath,
                amssymb,
                showpacs,
                superscriptaddress]{revtex4-1}
\usepackage{graphicx} % Include figure files
\usepackage{dcolumn}  % Align table columns on decimal point
\usepackage{bm}       % bold math
\usepackage{amsfonts}
\usepackage{dsfont}
\usepackage{csquotes}

\usepackage{url}
\usepackage{ulem}
\usepackage{color}

\usepackage{float}
\usepackage{physics}

\begin{document}

\title{High-harmonic spectroscopy of light-driven nonlinear anisotropic anomalous Hall effect in a Weyl semimetal}

\author{Amar Bharti}
\affiliation{%
Department of Physics, Indian Institute of Technology Bombay,
           Powai, Mumbai 400076, India }

\author{M. S. Mrudul}
\affiliation{%
Department of Physics, Indian Institute of Technology Bombay,
           Powai, Mumbai 400076, India }
                     
\author{Gopal Dixit}
\email[]{gdixit@phy.iitb.ac.in}
\affiliation{%
Department of Physics, Indian Institute of Technology Bombay, Powai, Mumbai 400076, India }

\date{\today}

%%%%%%%%%%%%%%%%% END OF PREAMBLE %%%%%%%%%%%%%%%%

\begin{abstract}
Weyl semimetals are promising quantum materials that offer unique topological properties. Lately, it has been shown that laser-driven electron dynamics have characteristic signatures in two-dimensional and three-dimensional Dirac semimetals. The transition from Dirac to Weyl semimetal requires the breaking of either inversion or time-reversal symmetry. 
The present work shows that the laser-driven electron dynamics in a 
Weyl semimetal with broken time-reversal symmetry has intriguing features in its high-harmonic spectrum. 
It is found that the parity and magnitude of the non-zero Berry curvature's components control  
the direction and strength of the anomalous current, which leads to the generation of the 
anomalous odd harmonics. 
We demonstrate  that the non-trivial topology of the Berry curvature in time-reversal symmetry  broken quantum materials can be probed by measuring the polarisation  
of the emitted anomalous odd harmonics.
Our findings unequivocally illustrate that laser-driven electron dynamics leads to the generation of 
nonlinear anisotropic anomalous Hall effect in time-reversal symmetry broken quantum materials on an ultrafast timescale.  
\end{abstract}

\maketitle

%\section{Introduction}
Discoveries of topological materials, such as   topological insulators, Dirac and Weyl semimetals, 
have revolutionised contemporary physics~\cite{hasan2010colloquium,  armitage2018weyl}. 
Moreover, these materials   
hold promises for upcoming technologies based on quantum science and electronics~\cite{keimer2017physics, sirica2021shaking, tokura2017emergent}. 
One of the remarkable properties of these materials is the robustness of the electronic states against perturbations, which 
has catalysed a plethora of interesting phenomena~\cite{qi2011topological, moore2010birth, yan2017topological}. 
Methods based on light-matter interaction play a  pivotal  role in probing and understanding various  
exotic properties of these topological materials~\cite{basov2017towards, bao2021light, mciver2012control}. 

If one increases the intensity of the light significantly,  various interesting perturbative and non-perturbative nonlinear  processes occur in the matter.  High-harmonic generation (HHG) is one such nonlinear process in which 
radiation of integer multiples of the incident light's frequency is emitted~\cite{ghimire2011observation, ghimire2019}. 
Numerous static and dynamic properties of solids have been probed  by analysing 
the emitted radiation during HHG~\cite{luu2015extreme, mrudul2021high, schubert2014sub, mrudul2021light, mrudul2021controlling, hohenleutner2015real, zaks2012experimental, pattanayak2020influence, langer2018lightwave, mrudul2020high, neufeld2021light, pattanayak2019direct}.   
In recent years, topological materials have turned out to be the centre of attention for HHG~\cite{reimann2018subcycle,  bai2021high, bauer2018high, dantas2021nonperturbative}. 
It is experimentally found that the bulk  and the topological surface 
 play different 
roles during HHG from a topological insulator~\cite{schmid2021tunable}. 
The interplay of  the time-reversal symmetry protection and the spin-orbit coupling in a topological insulator leads to anomalous dependence of harmonic yield on the polarisation of the driving laser~\cite{baykusheva2021all}. 
Berry curvature plays an important role  in determining the behaviour of high-harmonic spectra in both  cases. 
In a three-dimensional Dirac semimetal, coherent dynamics of the Dirac electrons plays the central role in HHG ~\cite{kovalev2020non, cheng2020efficient}.  Moreover, it has been reported that the 
nonlinear responses of the three- and two-dimensional Dirac semimetals are significantly different~\cite{lim2020efficient}.  
In all cases, time-reversal symmetry (TRS) is inherently preserved in topological insulators and Dirac semimetals.  
Therefore, it is natural to envision exploring how the breaking of the TRS affects HHG from topological materials and is the main emphasis of the present work.

A Weyl semimetal  (WSM) can be formed either by breaking time-reversal or inversion symmetry of the corresponding Dirac semimetal phase, and either case results in non-zero Berry curvature~\cite{yan2017topological}.
A Weyl semimetal  consists of the topologically protected degenerate points,  known as 
Weyl points, which can be seen as the monopoles of the Berry curvatures in momentum space~\cite{armitage2018weyl}. 
This makes WSM  as one of the most exotic gapless systems. 
In 2015, the first WSM was realised experimentally in transition-metal monopnictides, which form the class of nonmagnetic 
WSM with broken inversion-symmetry~\cite{xu2015discovery, xu2015discovery1, lv2015experimental}. 
Later, three groups have shown  the evidence of magnetic WSM in ferromagnetic materials with broken TRS experimentally~\cite{morali2019fermi, liu2019magnetic, belopolski2019discovery}. 

Present work focuses on addressing some crucial questions such as how TRS breaking and resultant modifications in Berry curvature affect HHG,  the role of the form of the Berry curvature's components, and 
how the separations of the Weyl points  influence HHG in WSM. In the following, we will demonstrate that non-zero Berry curvature in TRS-broken WSM leads to anomalous current in a direction perpendicular to the electric field and anomalous odd harmonics -- analogous to the anomalous Hall effect. Moreover, we will show that the directions of the emitted anomalous odd harmonics are related to the nature of the 
Berry curvature's components. 
The appearance of the anomalous odd harmonics allows us to 
probe non-trivial topology of the TRS-broken WSM by measuring 
the polarisation of the emitted anomalous odd harmonics.
Recently,  HHG from an inversion-symmetry broken WSM was explored experimentally in which a linearly  polarised pulse leads to the generation of even harmonics, related to non-zero Berry curvature~\cite{lv2021high}.  
Our findings are in contrast to previously reported works where Berry curvature mediated  
anomalous electron's velocity leads to the generation of even harmonics~\cite{schubert2014sub, liu2017high, hohenleutner2015real, luu2018measurement}.

The Hamiltonian corresponding to WSM with broken TRS can be written as~\cite{hasan2017discovery}
 \begin{equation}
 \mathcal{H}(\mathbf{k}) = \bm{d}(\mathbf{k})\cdot \bm{\sigma} = 
 d_1(\mathbf{k})\sigma_x + d_2(\mathbf{k}) \sigma_y + d_3(\mathbf{k})\sigma_z. 
 \label{eq1} 
 \end{equation}
 Here, $\bm{\sigma}$ is the Pauli vector and 
 [$d_1 =t_x \{ \cos(k_x a) - \cos(k_0 a)\} + t_y \{\cos(k_y b) -1\} + t_z \{\cos(k_z c) -1\}$,
 $ d_2 = t_y \sin(k_y b)$ and $d_3 = t_z \sin(k_z c)$] with 
 $(\pm k_0,0,0)$ as the positions of  the Weyl points, 
 $t_{x,y,z}$ as hopping parameters and  $a, b, c$ are lattice parameters. 
Here, we assume ferromagnetic WSM with tetragonal crystal structure, i.e., $a = b \neq c$~\cite{meng2019large}.  
$a = b = 3.437~ \text{\AA}, c = 11.646 ~\text{\AA}$ and $t_{x} = 1.88~\textrm{eV}, t_{y} = 0.49~\textrm{eV}, 
t_{z} = 0.16~\textrm{eV}$ are considered. 
The parameters used here are in accordance with the ones used in Ref.~\cite{nematollahi2020topological}. 
It is easy to see that the above Hamiltonian exhibits TRS breaking, i.e., 
$\hat{\mathcal{T}}^\dagger {\mathcal{H}}(-\mathbf{k})\hat{\mathcal{T}}\neq  {\mathcal{H}}(\mathbf{k})$. 
However,  $\hat{\mathcal{P}}^\dagger {\mathcal{H}}(-\mathbf{k})\hat{\mathcal{P}} = {\mathcal{H}}(\mathbf{k})$ 
ensures that inversion symmetry is preserved. 
Here, $\hat{\mathcal{P}} = \sigma_{x}$ and $\hat{\mathcal{T}} = \hat{\mathcal{K}}$
such that $ \hat{\mathcal{K}}^{\dagger} i  \hat{\mathcal{K}} = -i$
are the inversion symmetry and time-reversal symmetry operators, respectively.
After diagonalizing the above Hamiltonian, 
energy dispersion can be obtained as $\mathcal{E}_\pm = \pm|\bm{d}(\mathbf{k})|=\pm \sqrt{d_1^2 + d_2^2 + d_3^2}$. We have considered $k_0$ = 0.2 rad/au.
The corresponding band-structure on $k_z=0$ plane is presented in Fig. S1~\cite{NoteX}.

Interaction of the laser with the WSM is modelled using semiconductor-Bloch equations in the  Houston basis as discussed in Refs.~\cite{mrudul2021light,floss2018ab}. 
Within this formalism, current at any time can be written as 
\begin{equation}
\mathbf{J}(\textbf{k}, t)= \sum_{m,n}\rho_{mn}^{\mathbf{k}}(t) \mathbf{p}_{mn}^{\mathbf{k_t}}, 
\label{eq:sbecurrent}
\end{equation} 
where $\mathbf{k_t} = \mathbf{k} + \mathbf{A(t)}$ with $ \mathbf{A(t)}$ as the vector potential of the laser, and
$\rho_{mn}^{\mathbf{k}}(t)$ is the density matrix at  time $t$. 
Here, $\textbf{A}(t)$ is related to its electric field $\textbf{E}(t)$ as $\textbf{E}(t) = -\partial \textbf{A}(t)/\partial t$.
$\mathbf{p}^{\mathbf{k_t}}_{nm}$ is group velocity matrix element and calculated as $\mathbf{p}^{\mathbf{k_t}}_{nm}= \matrixel{n,\mathbf{k_t}}{\grad_{\mathbf{k_t}}\mathcal{H}_{\mathbf{k_t}}}{m,\mathbf{k_t}}$.
By performing the integral over entire Brillouin zone and taking Fourier transform ($\mathcal{FT}$), high-harmonic 
spectrum is simulated as 
\begin{equation}
\mathcal{I}(\omega) = \left|\mathcal{FT}\left(\frac{d}{dt} \left[\int_{BZ} \textbf{J}(\mathbf{k}, t)~d{\textbf{k}} 
\right] \right) \right|^2.
\end{equation}

\begin{figure}[h!]
\includegraphics[width=  \linewidth]{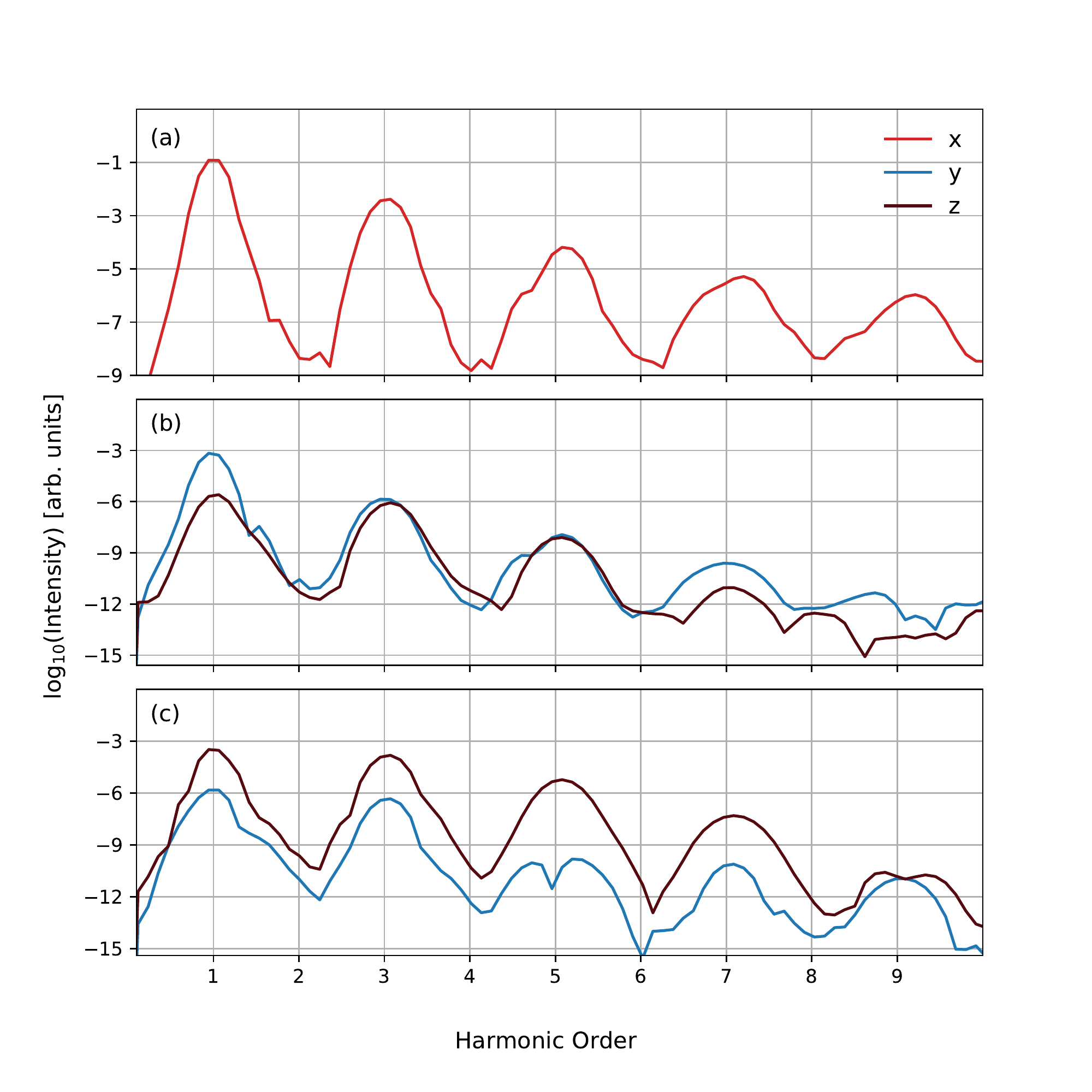}
\caption{High-harmonic spectra corresponding to a linearly  polarised pulse. The pulse is polarised  along 
(a) $x$, (b) $y$, and (c) $z$ directions. 
The driving pulse is $\simeq$ 100 fs long with intensity $1 \times 10^{11}$ W/cm$^2$, and wavelength 3.2 $\mu$m.  Decoherence time of 1.5 fs is added phenomenologically in semiconductor Bloch equations.}  \label{fig2}
\end{figure}

Figure~\ref{fig2} presents high-harmonic spectra corresponding to linearly polarised pulse.
When the pulse is polarised along the $x$ direction, odd harmonics are generated  along the laser polarisation  as evident from Fig.~\ref{fig2}(a). 
However, results become intriguing  when the pulse is  polarised along the $y$ or $z$ direction. 
In both  cases, odd harmonics are generated along the laser polarisation. 
Moreover, anomalous odd harmonics  along perpendicular directions are also generated.  
As reflected from Figs.~\ref{fig2}(b) and (c), when laser is polarised 
along the $y$ or $z$ direction, anomalous odd harmonics along  
the $z$ or $y$ direction, respectively, are generated.  
However, the yield of the anomalous  harmonics is relatively weaker  in comparison to the 
parallel  harmonics. 
On the other hand,  odd and even harmonics 
are generated in an  inversion-symmetry broken 
WSM~\cite{lv2021high}.
Furthermore, it has been 
concluded that  the appearance of the even harmonics is related to the 
spike-like Berry curvatures in  inversion-symmetry broken WSM~\cite{lv2021high}.

To understand why a linearly polarised driving  pulse leads to parallel and anomalous odd harmonics, and how these 
findings are related to TRS breaking in WSM, we employ  
the semiclassical equation  
of Bloch electrons in an external electric field $\textbf{E}(t)$. Within this approach, expression of the anomalous current is written as
$\bm{J}_\Omega(t) = -\textbf{E}(t) \times  \int \boldsymbol{\Omega}_\mu(\mathbf{k})~\rho_{ \mu}(\mathbf{k},t)~ d \mathbf{k}$ with 
 $\boldsymbol{\Omega}_\mu$ and $\rho_\mu(\mathbf{k},t)$ as the Berry curvature and band-population of the $\mu^{th}$ energy band, respectively~\cite{liu2017high}. 
We can assume that the initial band population is symmetric under inversion as 
$\rho(\mathbf{k}, 0) = \rho(-\mathbf{k}, 0)$. 
In the presence of a laser, momentum of an electron changes 
from $\mathbf{k}$ to $\mathbf{k_t}$, which leads 
to the change in the  band population as $ \rho(\mathbf{k}, t) = \rho(\mathbf{k_t},0)$. 
Under time-translation of the laser $t\rightarrow t+T/2$, the anomalous current can be expressed as 
\begin{eqnarray}
\bm{J}_\Omega(t+T/2)  & = & -\mathbf{E}(t+T/2) \times \int \boldsymbol{\Omega}_\mu(\mathbf{k})~\rho_\mu\big(\mathbf{k}_{\mathbf{t}+T/2},0\big) ~d \mathbf{k} \nonumber \\
    & = & \mathbf{E}(t) \times \int\boldsymbol{\Omega}_\mu(\mathbf{k}) ~\rho_\mu(\mathbf{k}-\mathbf{A}(t), 0\big) ~d \mathbf{k}  \nonumber \\
   & = &  \mathbf{E}(t) \times  \int \boldsymbol{\Omega}_\mu(\mathbf{k}) ~\rho_\mu\big(\mathbf{k_t}, 0\big) ~d \mathbf{k} \nonumber \\
   & = &  - \bm{J}_\Omega(t). 
\end{eqnarray}
In the above equations, we have used $\mathbf{E}(t+T/2) = - \mathbf{E}(t)$,  
$ \mathbf{A}(t+T/2) = -\mathbf{A}(t)$,  and $\rho(\mathbf{k}, 0) = \rho(-\mathbf{k}, 0)$; 
and changed the dummy variable  $\mathbf{k}\rightarrow-\mathbf{k}$ in the integral. 
Also, Berry curvature for an inversion-symmetric system with  broken TRS  obeys  
$ \boldsymbol{\Omega}(\mathbf{k})= \boldsymbol{\Omega}(-\mathbf{k})$.

The contribution of $\bm{J}_\Omega (t)$ to the $n^{\textrm{th}}$ harmonic is given by 
$  \bm{J}^n_\Omega(\omega) \propto \int_{-\infty}^\infty \bm{J}_\Omega(t) e^{in\omega t}~dt$. 
By changing  $t \rightarrow t + T/2$ in the integral, we obtain  
$ \bm{J}^n_\Omega(\omega) \propto  \int_{-\infty}^\infty \bm{J}_\Omega(t+T/2) e^{in\omega (t+T/2)}~ dt = 
-e^{in\pi}\int_{-\infty}^\infty \bm{J}_\Omega(t) e^{in\omega t}~dt$, which implies that only odd harmonics are allowed 
as $\exp(in\pi) = -1$. 
Thus, TRS-broken systems lead to anomalous odd harmonics, 
which is in contrast to the case of TRS-preserving systems with broken-inversion symmetry in which the anomalous current leads to the generation of even harmonics~\cite{schubert2014sub, liu2017high, hohenleutner2015real, luu2018measurement}. 

In order to discern the directions of the  anomalous current, we need to understand the distinct   
role  of the Berry curvature's components, which is written as 
$\boldsymbol{\Omega}(\mathbf{k}) = \Omega_{k_{x}}(\mathbf{k}) \hat{e}_{k_{x}} + 
\Omega_{k_{y}}(\mathbf{k}) \hat{e}_{k_{y}} + \Omega_{k_{z}}(\mathbf{k}) \hat{e}_{k_{z}}$. 
The  expressions of the Berry curvature's  components corresponding to the Hamiltonian in Eq.~(\ref{eq1}) are given in the supplementary material (see Eqs. S2-S4~\cite{NoteX}). 
The direction of the anomalous current is given by $\mathbf{E} \times \mathbf{\Omega}$ and the integral is perform over the entire Brillouin zone. 
Moreover, $\mathbf{E}$ is a function of time and $\boldsymbol{\Omega}$ is function of $\mathbf{k}$, so their product does 
not changes the parity. 

\begin{figure}[h!]
\includegraphics[width= 1.05 \linewidth]{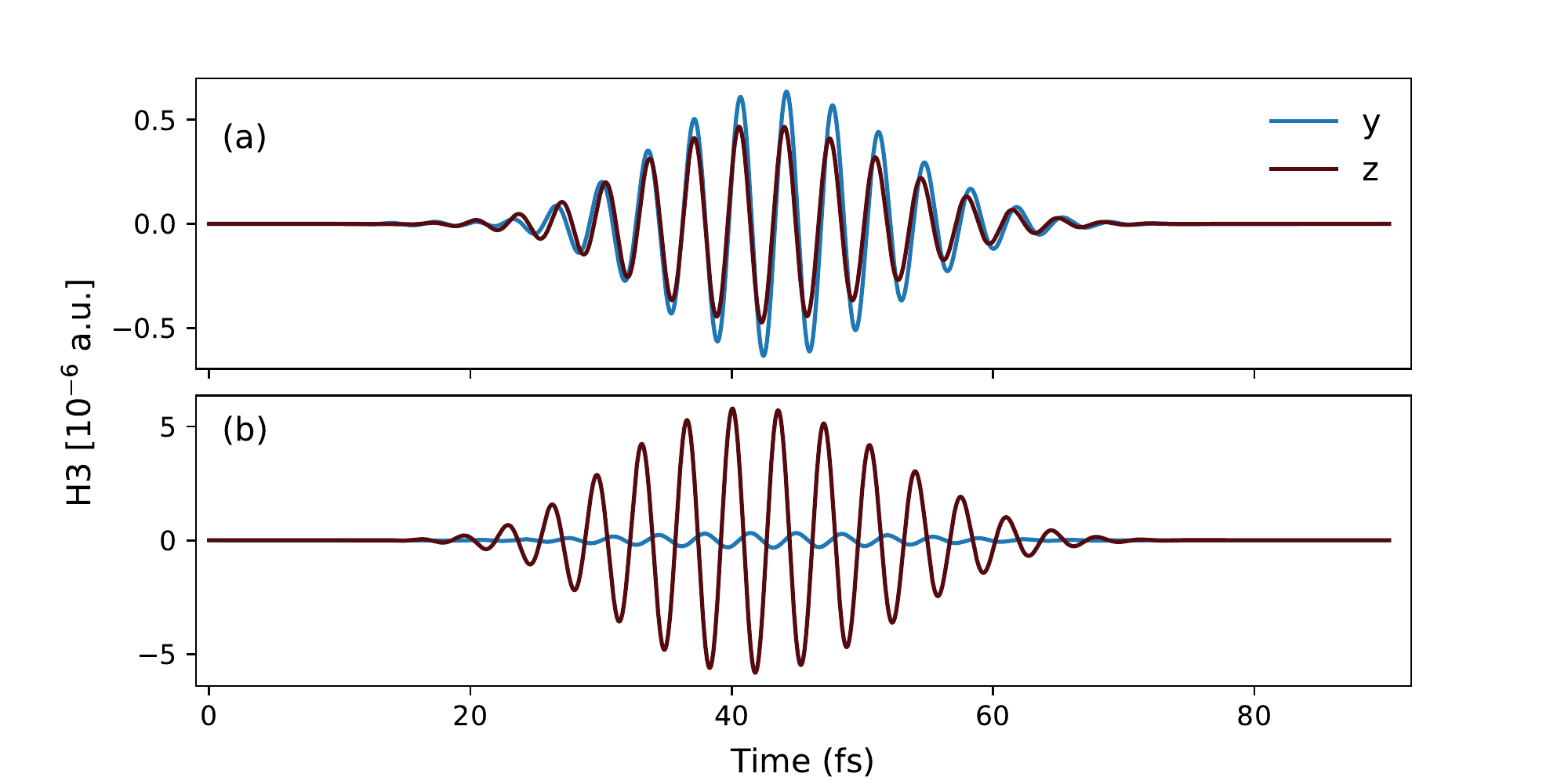}
\caption{Third harmonic (H3) in the time domain. 
Driving laser pulse is linearly polarised along (a) $y$  and (b) $z$ directions. The driving pulse has the same parameters as in Fig.~\ref{fig2}.}  \label{fig3}
\end{figure} 

If the laser is polarised along the $x$ direction, then it is straightforward to see that the anomalous current along 
the $y$ and $z$ directions turn out to be zero as $\Omega_{k_{y}}$ and $\Omega_{k_{z}}$ are odd functions in the two directions. 
On the other hand, $\Omega_{k_{x}}$ is an even function in all  directions, contributing to the anomalous current when the laser is polarised along $y$ or $z$ direction. 
Thus, present theoretical analysis is consistent with numerical results shown in Fig.~\ref{fig2}, which unequivocally establishes that non-trivial topology of the Berry curvature leads to nonlinear anomalous odd harmonics -- the light-driven 
nonlinear anomalous Hall effect.

At this point it is natural to investigate what determines the phase between the 
parallel  and the anomalous harmonics. To address this issue, we focus on 
the third harmonic (H3) in the time domain.
When the laser is polarised along $y$ direction, H3 along the $y$ and $z$ directions is in phase as evident from  Fig.~\ref{fig3}(a). However, it becomes out of phase in the case of  the $z$ polarised pulse.  The reason behind the in phase or out-of-phase of H3 can be attributed to the sign of 
$\bm{J}_\Omega (t)  \propto \int \mathbf{E}(t) \times \mathbf{\Omega}(\mathbf{k}) d\mathbf{k}$, which yields positive (negative) sign when laser is along the $y   (z)$ direction. 

\begin{figure}[h!]
\includegraphics[width=  \linewidth]{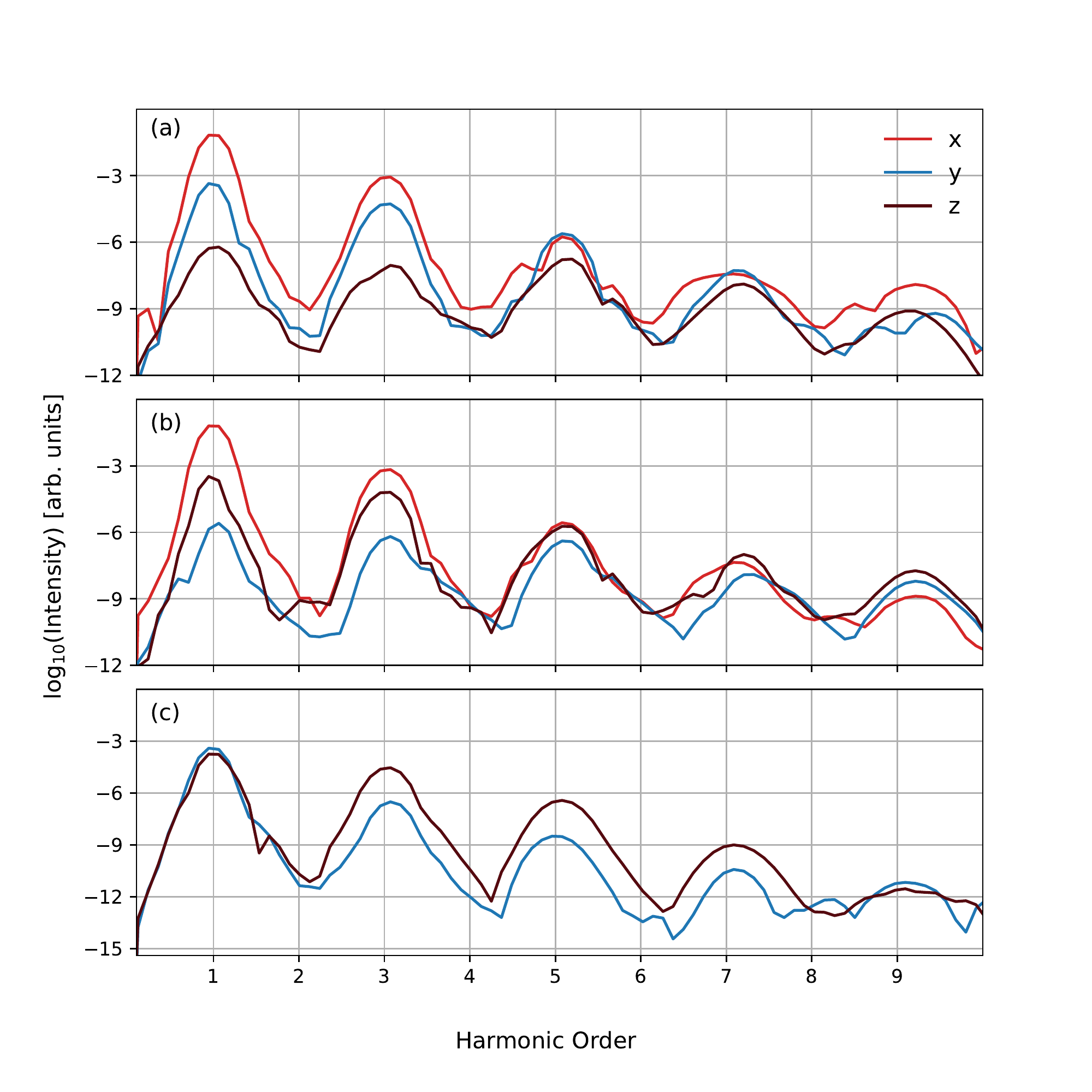}
\caption{High-harmonic spectra generated by the right-handed circularly polarised pulse in 
(a) $x-y$, (b) $x-z$, and (c) $y-z$ planes. The parameters of the laser and decoherence time are the same as given in 
Fig.~\ref{fig2}.}  \label{fig4}
\end{figure} 

To corroborate  our findings about the generation of the  anomalous odd 
harmonics and its relation with non-trivial topology of 
the Berry curvature's component, high-harmonic spectra generated by circularly polarised pulse 
are presented in Fig.~\ref{fig4}. 
In agreement with the two-fold rotation symmetry of the Hamiltonian, only odd harmonics are generated.
When the pulse is on $x-y$ plane, odd harmonics along the $x$ and $y$ directions are generated as 
$x$ and $y$ components of the 
driving electric field are non-zero.  
Moreover,  due to the non-zero $y$ component of the driving field, 
anomalous odd harmonics are generated along the $z$ direction 
[see Fig.~\ref{fig3}(a)]. In this case,  the mechanism   
is same as it was in the case of the linearly polarised pulse along the $y$ direction. 
The same is applicable in the case of the circularly polarised  pulse on $x-z$ plane. In this case, 
parallel odd harmonics are generated along the $x$ and $z$ directions, whereas anomalous odd harmonics are generated along the $y$ direction [see Fig.~\ref{fig3}(b)]. However, when the pulse is polarised on $y-z$ plane, only parallel harmonics along the $y$ and $z$ directions are generated, and no anomalous harmonics along the $x$ direction  are generated. This is expected due to even and odd natures of   
$ \Omega_{k_{x}}$ and $ \Omega_{k_{y}/k_{z}}$, respectively (see Eqs. S2-S4~\cite{NoteX}).

\begin{figure}[h!]
\includegraphics[width=  \linewidth]{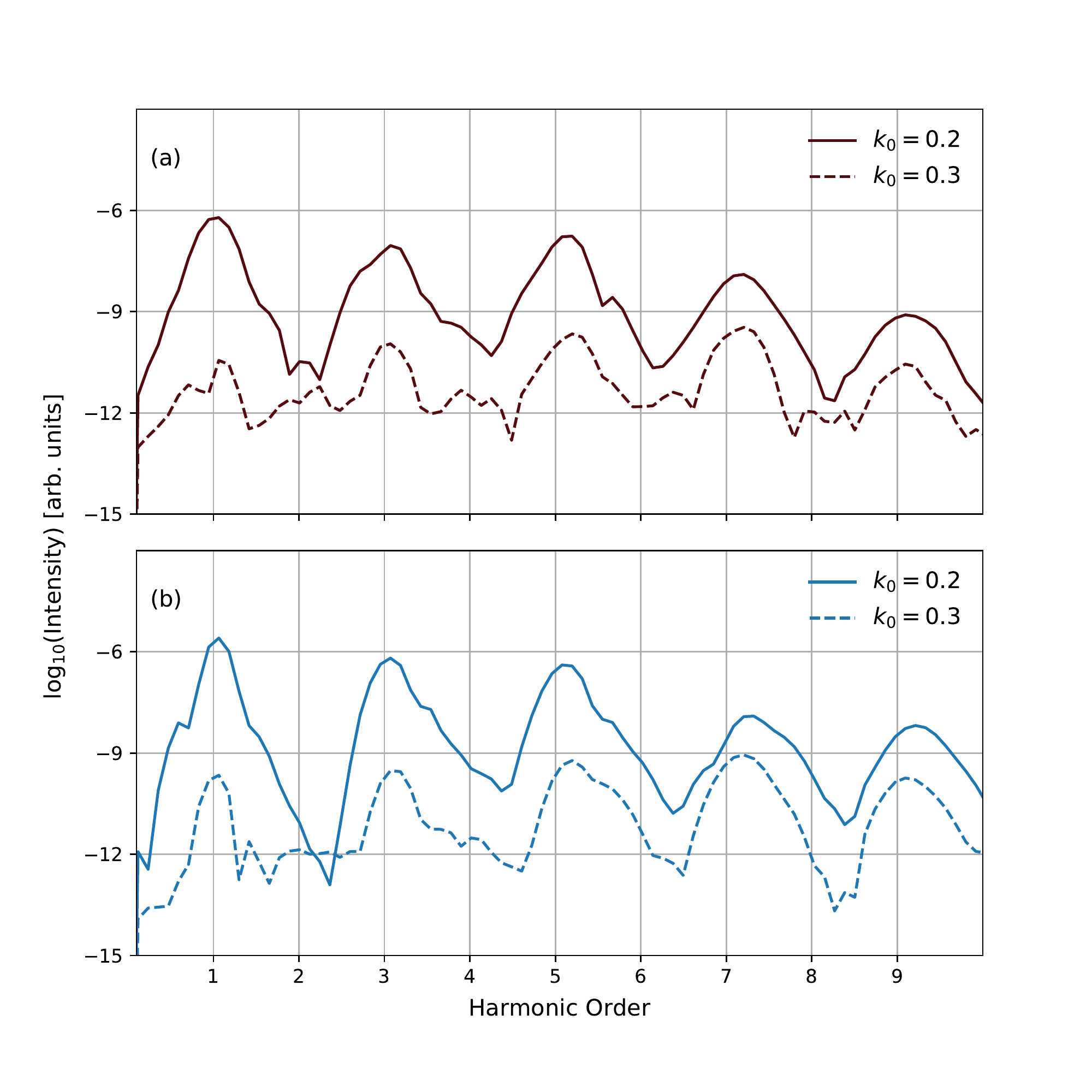}
\caption{Comparison of the anomalous high-harmonic yield for different values of $k_0$ = 0.2 and 0.3 rad/au. 
Odd anomalous harmonics along (a) the $z$ direction when polarisation of the pulse is on $x-y$ plane, and  (b) 
the $y$ direction when polarisation of the pulse is on $x-z$ plane. 
$k_0$ is proportional to the distance between the two Weyl points.}  \label{fig5}
\end{figure} 

After establishing the non-trivial role of the 
Berry curvature's components  and their parity,  
let us explore  how their strengths affect the yield of the anomalous odd harmonics. 
We know that  the magnitude of the  anomalous current depends on $\mathbf{E}\times \mathbf{\Omega}$. Moreover, the magnitude of the Berry curvature's components depend on $k_0$ (see Fig. S2~\cite{NoteX}).
Therefore, as we change the value of  $k_0$ from $0.2$ to $0.3$, the strength of the Berry curvature's components reduces, which lead to the reduction in the strength of the anomalous current. 
Fig.~\ref{fig5}  presents a comparison of the  yield of the anomalous harmonics  for two different values of 
$k_0$. 
In the case of circularly polarised pulse on $x-y$ plane, the anomalous harmonics  along the $z$ direction reduces drastically  as we change $k_0$ from $0.2$ to $ 0.3$ rad/au [see Fig.~\ref{fig5}(a)]. 
The same is true for the pulse on the $x-z$ plane and anomalous harmonics  along the 
$y$ direction [see Fig.~\ref{fig5}(b)].   
Therefore, the yield of the anomalous harmonics reduces drastically as 
 the value of $k_0$ increased from 0.2  to 0.3 rad/au. 
Similar conclusions can be drawn in the case of HHG from the linearly polarised pulse (see Fig. S3~\cite{NoteX}). 
However, the yield of the parallel harmonics is insensitive to the change in the value of $k_0$  
 (see Figs. S4 and S5~\cite{NoteX}).
Our findings are similar to the anisotropic anomalous Hall effect in which the magnitude of the current depends on the integral of the Berry curvature~\cite{yang2021noncollinear}.

Not only the anomalous current and harmonics encode the non-trivial symmetry and the magnitude of the Berry curvature's components but also  the anomalous current and harmonics tailor the polarisation of the emitted harmonics, which offer an elegant way  to probe non-trivial topological properties of the Berry curvature by an all-optical way. 
As evident from Fig.~\ref{fig3},  $y$ and $z$ components of H3 are in-phase 
and out-of-phase when the driving laser is polarised along the $y$ and $z$ directions, respectively, which gives two different polarisation of H3 (see Fig. S6(a)~\cite{NoteX}). 
Thus, by measuring the polarisation of H3, 
the non-trivial topology of the Berry curvature can be probed as it controls 
the strength and the phase between  
$y$ and $z$ components of H3. 
The same observations are true for other higher-order  harmonics corresponding to  linearly polarised driver (see Figs.~\ref{fig2} and S6~\cite{NoteX}). 
Similar conclusions can be made when circularly polarised laser is used for HHG (see Figs.~\ref{fig4} and S7~\cite{NoteX}).

In summary, we have investigated  the role of TRS breaking in the strong-field driven 
nonlinear process in topological materials.  
For this purpose, the inversion-symmetric Weyl semimetal with broken TRS is considered.   
It is found that the 
non-trivial topology of the TRS-broken Weyl semimetal leads to the generation  of the anomalous odd harmonics, which are anisotropic and appear only when the driving laser has non-zero components along the $y$ or $z$ direction. 
Non-trivial symmetry of the Berry curvature's components of the 
TRS-broken Weyl semimetal is responsible for 
the anisotropic nature of the anomalous harmonics (current). 
Moreover, the strength of the Berry curvature dictates 
 the strength of the anomalous odd harmonics. 
Furthermore, non-trivial topology properties of the Berry curvature and its strength 
can be probed by measuring the the polarisation of the emitted anomalous odd harmonics.
Present  work opens a new avenue for studying strong-field driven electron dynamics and high-harmonic generation in systems with broken TRS, such as exotic magnetic and topological materials; and tailoring the polarisation of the emitted harmonics. 

G. D. acknowledges useful discussion with Prof. Sumiran Pujari (IIT Bombay). 
G. D. acknowledges support from Science and Engineering Research Board (SERB) India 
(Project No. ECR/2017/001460) and the Ramanujan fellowship (SB/S2/ RJN-152/2015). 

%\newpage
%\bibliography{solid_HHG}

%merlin.mbs apsrev4-1.bst 2010-07-25 4.21a (PWD, AO, DPC) hacked
%Control: key (0)
%Control: author (8) initials jnrlst
%Control: editor formatted (1) identically to author
%Control: production of article title (-1) disabled
%Control: page (0) single
%Control: year (1) truncated
%Control: production of eprint (0) enabled
%

\end{document}